\documentclass[%
 reprint,
nofootinbib,
 amsmath,amssymb,
 aps,
]{revtex4-2}

\usepackage{graphicx}
\usepackage{dcolumn}
\usepackage{bm}
\usepackage[]{hyperref}
\usepackage{amsmath}
\usepackage{mathrsfs}
\usepackage{gensymb}
\usepackage{tensor}
\usepackage{physics}
\usepackage{amsfonts}
\usepackage{amssymb}
\usepackage{comment}
\usepackage{subfigure}
\usepackage{xcolor}
\usepackage{soul}
\usepackage[normalem]{ulem}

\begin{document}

\preprint{APS/123-QED}

\title{Geodetic precession and shadow of quantum extended black holes}

\author{Reza Saadati}%
\author{Fatimah Shojai}
\email{fshojai@ut.ac.ir}
\affiliation{%
	Department of Physics, University of Tehran,\\Kargar North Street, Tehran, Iran 14395/547 
}%

\date{\today}

\begin{abstract}
 We study the circular motion of massive and massless particles in a recently proposed quantum-corrected Schwarzschild black hole in loop quantum gravity. This solution is supposed to introduce small but non-zero quantum corrections in the low curvature limit. In this paper, we confine our attention to the shadow of the black hole and the geodetic precession of
 a freely falling gyroscope in a circular orbit. Despite the mathematical complexity of the metric, our results are exact and show that the black hole shadow decreases slightly in this solution while the quantum corrections introduce a non-trivial term in the geodetic precession frequency of the gyroscope.

\end{abstract}

\maketitle

\section{Introduction}
General relativity (GR), as a theory of gravity, uses the geometry of the space-time continuum to describe the gravitational potential. However, this theory breaks down at genuine singularities of space-time. This is because the geometry becomes ill-defined in the vicinity of such singularities. Solving this problem has motivated many efforts either to construct some extended theories of GR or to formulate a consistent theory of quantum gravity.

In particular, in space-time containing a singular black hole (BH), there are two approaches to deal with this problem. The first, which we are not interested in here, imposes some exotic conditions on the matter fields \cite{Zaslavskii:2010qz} to avoid the BH singularity. In particular, violation of the strong energy condition, which is one of the basic assumptions of the Hawking–Penrose singularity theorems \cite{Penrose:1964wq,Hawking:1970zqf} could lead to a non-singular BH as the final state of gravitational collapse. For a thorough review of this approach see \cite{Ansoldi:2008jw}. Moreover, for the formation of a particular class of regular BHs from the gravitational collapse of a massive star we refer the reader to \cite{Shojai:2022pdq}.

The second approach involves quantum effects. It is  well known that at scales smaller than the Planck length, quantum effects come into play. Thus, quantum effects become important to construct a consistent theory of gravity in this regime. Loop quantum gravity (LQG) emerges as a possible theory of quantum gravity by quantizing the geometry of space-time \cite{Ashtekar:2013hs,Rovelli:1997yv}. Roughly speaking, in LQG one defines the area operator whose smallest non-zero eigenvalue $\Delta$ introduces some modifications to Einstein's equations. At $\Delta=0$, one recovers the singular results of GR while its strictly positive value in LQG is the key to avoid curvature singularities \cite{Singh:2009mz,Ashtekar:2011ni}. LQG has been successfully applied to homogeneous cosmology where at high curvature regime, i.e. Planck scale, the quantum geometry modification dominates and removes the Big Bang singularity. In the low curvature regime, loop quantum cosmology (LQC) approximately reproduces the results of Einstein field equations, see \cite{Ashtekar:2011ni} and references therein for more details. The lesson learned from LQC, led the community to invoke LQG to resolve the singularity of the Schwarzschild BH. The desired solution should satisfy two conditions. First, the physical effects should be insensitive to the classical phase space structure, and second, at low curvature scales, the quantum effects of the geometry should be negligible. For example, the models presented in \cite{Boehmer:2007ket,Chiou:2008nm,Corichi:2015xia,Olmedo:2017lvt,Khodadi:2020jij,Bouhmadi-Lopez:2019hpp,Ashtekar:2005qt,Modesto:2005zm,Cartin:2006yv,Campiglia:2007pb} suffer from the violation of one of these conditions.

Recently Ashtekar, Olmedo, and Singh (AOS) proposed a regular static spherically symmetric BH solution of LQG \cite{Ashtekar:2018cay,Ashtekar:2018lag,Ashtekar:2020ckv} that is said to be consistent with the above conditions. In their solution, while quantum corrections  give rise to large violations of Einstein’s equations near the would-be singularity, they become  insignificant in the low-curvature regime, i.e., near and outside the horizons of macroscopic BHs \cite{Ashtekar:2018cay}. Moreover, the effective AOS solution includes both the interior and exterior regions between the horizon and infinity \cite{Ashtekar:2020ckv,Ashtekar:2018cay}.
By calculating the Paczynski–Wiita pseudo-Newtonian potential \cite{Paczynski:1979rz,Abramowicz:2009bh} for the AOS geometry, it has been shown that the quantum corrections to the innermost stable circular orbit (ISCO) diverge \cite{Faraoni:2020stz}. Therefore, the non-zero quantum parameter, no matter how small, could have a significant effect in the low curvature regime.

The main purpose of this paper is to study  two  different features of AOS spacetime including the geodesic precession and the BH shadow. For this purpose,  we consider circular orbits of free massive and massless particles in this space-time. Apart from the Lense-Thirring (LTP) precession, which is caused by the coupling of the angular momentum of the rotating central body with the spin of the gyroscope \cite{Lense:1918zz}, the coupling of the orbital angular momentum of the gyroscope with its spin leads to a precession which is known as the de Sitter effect or geodetic precession (GP) \cite{de Sitter:1916}. 
LTP and GP are two famous examples of gravitomagnetic effects \cite{Mashhoon:1984fj,Jantzen:1992rg,Maartens:1997fg,Clark:2000ff,Ruggiero:2002hz}. For a good review see \cite{Ruggiero:2002hz}. Motivated by the Gravity Probe B experiment \cite{Everitt:2011hp,Everitt:2015qri}, we are interested here in the GP effect in AOS space-time. Using a test gyroscope rotating in a circular orbit in the equatorial plane of the AOS geometry, one can analyze the motion of the gyroscope's spin vector and thus probe the geometry of spacetime. We then compare our results with the GP measurements from Gravity Probe B to see how quantum corrections affect these phenomena. For a detailed discussion of GP in modified theories of gravity, we refer the reader to \cite{Fathi:2020sey,Finch:2016bum,Farrugia:2020fcu,Matsuno:2009nz,Said:2013gna,Azreg-Ainou:2019ylk,Overduin:2013tma,Chen:2010yx}.

While the Gravity Probe B experiment was conducted in the weak-field regime, the strong gravity of the AOS BH metric can be reflected in the modification of the photon ring and the shape of the shadow \cite{Synge:1966okc,Luminet:1979nyg,Cunha:2018acu,Perlick:2021aok}. 
However
note that in this regime the spacetime curvature is still much lower than expected for regions near the central singularity. Therefore, we expect that quantum corrections cannot leave significant imprints in the shadow of macroscopic BHs. The calculation of the BH shadow is very important for observational purposes. In particular, due to the observation of the BH shadow in the center of the nearby radio galaxy $M87^*$ by the Event Horizon Telescope collaboration \cite{EventHorizonTelescope:2019dse}, the existence of BHs is now an undeniable fact. In this paper we try to calculate the shadow of the AOS BH exactly. We find that, as expected in the case of macroscopic BHs, the quantum effects on the BH shadow can be neglected. Moreover, our result is in agreement with the numerical results of \cite{Devi:2021ctm}.

The line element outside the horizon of the effective geometry of the quantum-corrected AOS BH with the metric signature $(-,+,+,+)$ is given by \cite{Ashtekar:2018cay,Ashtekar:2018lag,Ashtekar:2020ckv}
\begin{equation}\label{eq:metr}
	ds^2=-h(r)dt^2+\frac{dr^2}{f(r)}+r^2d\Omega^2
\end{equation}
in which
\begin{equation}\label{eq:f}
	f(r)=1-\left(\frac{2m}{r}\right)^{1+\epsilon},\quad 	h(r)=\left(\frac{r}{2m}\right)^{2\epsilon}f(r),
\end{equation}
$d\Omega^2=d\theta^2+\sin^2\theta d\phi^2$ is the line element on the unit $2-$sphere and $m$ is the ADM mass \cite{Ashtekar:2020ckv}. In \eqref{eq:f}, the dimensionless parameter $\epsilon$ gives the quantum corrections to Schwarzschild spacetime. For a macroscopic black hole, it approximately satisfies \cite{Ashtekar:2018lag}
\begin{equation}\label{eq:epsilon}
	\epsilon\approx \left(\frac{\gamma^2\Delta}{16\pi m^2}\right)^{1/3}
\end{equation}
where $\gamma\approx0.2375$ is the Barbero–Immirzi parameter and $\Delta$, the area gap, is related to the Planck length $l_p$ by $\Delta\approx5.17l_p^2$.  It is easy to see that the surface $r=2m$ is the event horizon for the AOS metric \eqref{eq:metr}. The asymptotic behavior of \eqref{eq:metr} is studied in \cite{Ashtekar:2020ckv}.  
 It is shown that despite of the fact that $g_{tt}$ diverges as $r\to \infty$, this geometry is actually asymptotically flat. By defining a new time co-ordinate $\bar{t}=t\left(\frac{r}{2m}\right)^{\epsilon}$, it can be shown that \eqref{eq:metr} approaches the Minkowski flat metric at spatial infinity \cite{Ashtekar:2020ckv}. Thus, one can define the asymptotically inertial frame that is fixed relative to the distant stars. Since $\epsilon$ is extremely small compared to unity, e.g., $\epsilon\approx10^{-26}$ for a solar mass BH, the metric components can be expanded as
\begin{equation}\label{eq:expanded1}
	-h(r)\approx -1+\frac{2m}{r}+2\epsilon\left(1-\frac{m}{r}\right)\ln\frac{2m}{r}
\end{equation}
\begin{equation}\label{eq:expanded2}
	f^{-1}(r)\approx \left(1-\frac{2m}{r}\right)^{-1}+2\epsilon \frac{m}{r}\left(1-\frac{2m}{r}\right)^{-2}\ln\frac{2m}{r}.
\end{equation} 
It is seen that the leading order quantum corrections to the Schwarzschild BH are represented by logarithmic terms. Therefore, for sufficiently large distances, it is more convenient to use the original form of \eqref{eq:metr} to avoid the logarithmic terms that would complicate the subsequent calculations. Moreover, it has been shown in \cite{Daghigh:2020fmw} that the metric \eqref{eq:f} does not provide a self-consistent Taylor approximation around $\epsilon=0$. Therefore, in order to find the main results of the paper, we avoid the use of Taylor expansions in all subsequent calculations. However, in order to see the quantum effect on quantities such as the innermost stable circular orbit, we expand our exact results up to the first order of $\epsilon$. 
Calculating the Ricci and Kretschmann scalars for the line element \eqref{eq:metr} yields
	\begin{equation}
		\mathcal{R}=-\frac{2\epsilon}{r^2}\left[1+\left(\frac{2 m}{r}\right)^{\epsilon +1}+\epsilon\right]
	\end{equation}
	\begin{multline}
		\mathcal{K}=\frac{4}{r^4}\Bigg[2 \epsilon  \left(\epsilon ^2-3 \epsilon +2\right) \left(\frac{2 m}{r}\right)^{\epsilon +1}+\epsilon ^2 \left(\epsilon ^2-2 \epsilon +3\right)\\
		+\left(2 \epsilon ^2-2 \epsilon +3\right) \left(\frac{2 m}{r}\right)^{2 \epsilon +2}\Bigg]
	\end{multline}
A direct inspection shows that the above quantities are well defined everywhere except at  $r\rightarrow0$.
Thus, as for the Schwarzschild solution, the AOS spacetime contains a curvature singularity hidden by a horizon.

This paper is organized as follows: In Section \ref{sec:Circular Orbits}, we obtain the effective potential for both time-like and null particles moving freely in the AOS BH background. Then, in Sections \ref{sec:TIME-LIKE CIRCULAR ORBITS} and \ref{Null CIRCULAR ORBITS}, we use the effective potential to study time-like and null circular orbits respectively. We then assume that a gyroscope rotates along time-like orbits and calculate the quantum modification of the GP frequency in Section \ref{sec:Geodetic Precession}. Finally, in Section \ref{se:CONCLUDING REMARKS} we review the highlights of the paper. Throughout this paper, we use geometrized units where $c=G=1$ and the Greek indices $\mu,\nu,...$ are used to denote the four-dimensional space-time described by the metric components $g_{\mu\nu}$.
\section{Effective Potential}\label{sec:Circular Orbits}
In this section, we are interested in time-like and null geodesics of metric \eqref{eq:metr}. Recall that the general motion of a particle in any spherically symmetric space-time can be reduced to the one-dimensional radial motion in an  effective potential. The particle Lagrangian can be written as
\begin{equation}\label{eq:first integral}
	2\mathcal{L}=\alpha=-h(r)\dot{t}^2+f^{-1}(r)\dot{r}^2+r^2\dot{\phi}^2
\end{equation}
where $\alpha=-1,0$ for time-like and null geodesics respectively  and dot denotes the derivative with respect to the particle's proper time. The coordinates are chosen such that the motion is restricted to the equatorial plane due to the conservation of angular momentum. Using \eqref{eq:first integral}, one can easily obtain the conserved energy and angular momentum of the particle as follows
\begin{equation}\label{eq:pt}
E=h(r)\dot{t},\hspace{1cm}L=r^2\dot{\phi}=r^2\omega
\end{equation}
where $\omega$ is defined as the proper angular velocity of the particle. These relations help us to express $\dot{t}$ and $\dot{\phi}$ in terms of the conserved quantities. The result is then substituted into \eqref{eq:first integral} to obtain a first order differential equation for the radial coordinate of the particle
\begin{equation}\label{eq:eomI}
	\frac{1}{2}\dot{r}^2+V_{\text{eff}}(r)=0
\end{equation}
where
\begin{equation}\label{eq:effective}
	V_{\text{eff}}(r)=\frac{1}{2}\left(\frac{L^2}{r^2}-\frac{E^2}{h(r)}-\alpha\right)f(r).
\end{equation}
It is clear from \eqref{eq:eomI} and \eqref{eq:effective} that the constant $E$ is included in the definition of the effective potential. Up to some irrelevant constants, \eqref{eq:eomI} and \eqref{eq:effective} are in agreement with \cite{Cardoso:2008bp}. For more details on circular orbits in static and dynamical space-time we refer the reader to \cite{Song:2021ziq}. In the following discussions we will make use of \eqref{eq:effective} to examine circular orbits in detail.

\subsection{TIME-LIKE CIRCULAR ORBITS}\label{sec:TIME-LIKE CIRCULAR ORBITS}
For a circular orbit 
$\dot{r}=0$,
equation \eqref{eq:eomI} leads to
\begin{equation}\label{eq:circulrOrbits}
	V(R)=0, \quad V'(R)=0.
\end{equation}
where $R$ is the radius of the orbit and prime denotes the derivative with respect to the radial coordinate. This allows us to identify the constants  $L$, $E$ of the test particle as follows\cite{Cardoso:2008bp}
\begin{align}
	\label{L:exact}
	L^2&=\frac{R^3h'(R)}{2h(R)-Rh'(R)} \nonumber\\
	&=\frac{2\epsilon-(\epsilon -1) \left(\frac{2 m}{R}\right)^{\epsilon +1}}{(\epsilon -3) \left(\frac{2 m}{R}\right)^{\epsilon +1}-2 (\epsilon -1)}R^2
\end{align}
\begin{align}\label{eq:E}
	E^2&=\frac{2h^2(R)}{2h(R)-Rh'(R)} \nonumber\\
	&=\frac{2}{R}\frac{\left(1-\left(\frac{2 m}{R}\right)^{\epsilon +1}\right)^2}{(\epsilon -3) \left(\frac{2 m}{R}\right)^{\epsilon +1}-2 (\epsilon -1)}\left(\frac{R}{2 m}\right)^{2 \epsilon }.
\end{align}
Assuming $R\gg 2m$, it is instructive to expand \eqref{eq:E} up to the first order in $\epsilon$ to obtain
\begin{equation}\label{eq:EE}
	E^2\approx 1+\epsilon\left(1+2 \ln \frac{R}{2 m}\right).
\end{equation}
The first term on the RHS of  \eqref{eq:EE} is the normalized particle's energy as defined in GR and the second one is the  correction due to the quantum space-time \eqref{eq:f}. It diverges as $R\rightarrow \infty$. 
Note that, for example, for a solar mass BH  with $m\sim 1.5\text{km}$, the quantum parameter is found to be $\epsilon=10^{-26}$ from \eqref{eq:epsilon}, and thus the quantum term in \eqref{eq:EE} can be neglected at the radius of the observable universe which is $R\sim 28.5\text{Gpc}\approx 10^{24}\text{km}$.

From \eqref{L:exact} and \eqref{eq:E}, one can find out how the quantum parameter $\epsilon$ affects the radius of the innermost circular orbit (ICO) by noting that $E$ and $L$ must be real-valued quantities. This implies that   
\begin{equation}\label{c1}
2h(R)-Rh'(R)>0
\end{equation}
which can be written as 
\begin{equation}\label{eq:Lc}
	(\epsilon -3) \left(\frac{2 m}{R}\right)^{\epsilon +1}-2 (\epsilon -1)>0
\end{equation} 
using \eqref{eq:f} \cite{Cardoso:2008bp}. We can easily see that for $\epsilon=0$, the above relation reduces to the well known result of GR, i.e. $R>3m$. 
From \ref{eq:Lc} one would expect that the ICO radius could be slightly modified by the quantum effects. To see this, we write \ref{eq:Lc} as
\begin{equation}\label{eq:ICO}
	R\geq R_{\text{ICO}}= 2m\left(\frac{\epsilon-3}{2\left(\epsilon-1\right)}\right)^{1/1+\epsilon}.
\end{equation}
This up to the first order in $\epsilon$ gives
\begin{equation}\label{co1}
	R_{\text{ICO}}\approx3m+ \left(\frac{\gamma ^2 m \Delta }{16 \pi }\right)^{1/3} \left(2-3 \ln\frac{3}{2}\right)
\end{equation}
where we have used \eqref{eq:epsilon}. This indicates that the radius of ICO in AOS space-time is larger 
than that in Schwarzschild space-time. Now, we turn our attention to the radius of ISCO 
and find the roots of $V''(R)=0$.  After some calculations, we can see that both equations $V''(R)=0$ and $dL/dR=0$ lead to a single algebraic equation as follows \cite{Cardoso:2008bp,Song:2021ziq}
\begin{equation}\label{eq:ISCOeq}
	h(R) \left[R h''(R)+3 h'(R)\right]-2 R h'^2(R)=0.
\end{equation}
Using \eqref{eq:f} with \eqref{eq:ISCOeq} and setting $\left(\frac{2m}{R}\right)^{1+\epsilon}\equiv x$, we obtain
a second-order algebraic equation for $x$
\begin{equation}\label{quad}
	-x^2 \left(\epsilon ^2-4 \epsilon +3\right)+x \left(3 \epsilon ^2-12 \epsilon +1\right)-4 (\epsilon -1)\epsilon=0
\end{equation}
with the following two exact solutions 
\begin{equation}\label{eq:Rpm}
	R_{\pm}=2m \left[\frac{A}{B\pm C}\right]^{1/1+\epsilon}
\end{equation}
where
\begin{equation}
	\begin{split}
		A&\equiv 2\left(\epsilon-3\right)\left(\epsilon-1\right),\quad B\equiv 3 \epsilon ^2-12 \epsilon +1\\
		C&\equiv (\epsilon +1) \sqrt{\epsilon  (22-7 \epsilon )+1}
	\end{split}.
\end{equation}
It is easy to show that $R_-$ is not real for $0<\epsilon\ll1$ while $R_+$ is real. Therefore, circular orbits whose radius is in the range $R_+<R<\infty$ are stable against radial perturbations. Hence, 
\begin{equation}\label{isco}
	R_{\text{ISCO}}=R_{+}.
\end{equation}
For $\epsilon=0$, equation \eqref{eq:Rpm} gives the well-known result of GR, $R_{\text{ISCO}}=6m$ and thus stable orbits exist in the range $6m<R<\infty$. One can expand $R_{\text{ISCO}}$ \eqref{eq:Rpm} up to the leading order of the quantum parameter $\epsilon$. This gives
\begin{equation}\label{eq:RISCO}
	R_{\text{ISCO}}\approx 6 m-\left(\frac{\gamma ^2 m \Delta }{2 \pi }\right)^{1/3}\left(4+3 \ln3\right).
\end{equation}
We can see from the above that $R_{\text{ISCO}}$ is smaller than the classical one. Substituting
$R_{\text{ISCO}}$ from \eqref{eq:Rpm} into \eqref{L:exact} and using \eqref{eq:f} lead to the minimum angular momentum 
\begin{equation}\label{eq:LMIN}
	L_{\text{min}}\approx2 \sqrt{3} m+  \left(\frac{\gamma ^2 m \Delta  }{2 \pi }\right)^{1/3} \frac{10-3 \ln 3}{\sqrt{3}}
\end{equation} 
up to the leading order of the $\epsilon$ expansion. This shows that up to the leading order, the minimum angular momentum of the particle is larger than that known for the Schwarzschild BH.
It is worth noting that the quantum correction to the ISCO radius \eqref{eq:RISCO} does not diverge as $\epsilon\to0$. 
This result contrasts sharply with \cite{Faraoni:2020stz}, which finds a divergent first-order quantum correction to $R_{\text{ISCO}}$ as $\epsilon\to0$. 
In \cite{Faraoni:2020stz}, the authors used $\epsilon$-expanded Paczynski-Wiita pseudo-Newtonian potential to find $R_{\text{ISCO}}$.  
Here, however, the standard method of the effective potential \cite{Cardoso:2008bp} was used and the result \eqref{isco} is obtained exactly, i.e. without using the Taylor expansion. For a  a clear comparison with \cite{Faraoni:2020stz}, in appendix \ref{appen}, we use the Paczynski-Wiita potential and recover the ISCO radius \eqref{isco}.

We now consider Kepler's third law in AOS spacetime. In Newtonian gravity, Kepler's third law states that the square of the orbital period of a particle is proportional to the cube of the length of the semi-major axis. The same holds for GR if we assume that the geometry of spacetime is described by the Schwarzschild line element \cite{Straumann:2013spu}.
Taking into account \eqref{eq:circulrOrbits}, the angular velocity $\Omega=d\phi/dt=\dot{\phi}/\dot{t}$ of the test particle is given by
	\begin{equation}\label{eq:Kep}
		\Omega^2=\frac{h'(R)}{2R}.
	\end{equation}
Note that $h'(r)>0$ for $\epsilon\ll1$ and thus $\Omega^2$ is a real function of $R$. Using \eqref{eq:f} and \eqref{eq:Kep}, we can derive the Kepler’s third law in AOS space-time
\begin{equation}\label{Kepler Law}
	\Omega^2R^3= \left[\epsilon  \left(\frac{R}{2 m}\right)^{2 \epsilon }-\frac{1}{2}\left(\epsilon-1\right)\left(\frac{R}{2 m}\right)^{\epsilon -1}\right]R
\end{equation}
For $\epsilon=0$ we obviously recover the Kepler's third law. 
To get some insight into the above relation, we expand it  up to the first order in $\epsilon$ 
\begin{equation}
	\Omega^2R^3\approx m+ \left(\frac{\gamma^2\Delta}{16\pi m^2}\right)^{1/3}\left(1-\frac{m}{R}\left[1+\ln\frac{2m}{R}\right]\right)R
\end{equation}
The second term on the RHS of the above relation  is positive  outside the event horizon and describes the LQG correction. Thus, for a given radius, the quantum effects lead to a slightly higher angular velocity compared to the result of GR.  In the limit $R\gg m$ we have
\begin{equation}
	\Omega^2R^3\approx m+\left(\frac{\gamma^2\Delta}{16\pi m^2}\right)^{1/3}R.
\end{equation}
In this way we obtain the remarkable result that the contribution of the quantum term to the particle period is proportional to the square of the orbital radius.\\
The angular velocity is a monotonically decreasing function of $R$. To see this, we take the derivative of the square root of \eqref{eq:Kep} with respect to $R$
\begin{equation}
	\frac{d\Omega}{dR}=\left(\frac{R}{2 m}\right)^{\epsilon }\frac{\left[4 \epsilon -(\epsilon -3) \left(\frac{2 m}{R}\right)^{\epsilon +1}\right](\epsilon -1) }{2 \sqrt{2} R^2 \sqrt{2 \epsilon -(\epsilon -1) \left(\frac{2 m}{R}\right)^{\epsilon +1}}}
\end{equation}
which is negative for $\epsilon\ll1$. Thus, it can be concluded that the angular velocity of the particle increases as $R$ decreases. Considering the  stable orbits, the angular velocity reaches its maximum value at $R=R_{\text{ISCO}}$ given by \eqref{isco}. Thus, up to the first order in $\epsilon$ one obtains
\begin{equation}
	m\Omega_{\text{ISCO}}\approx \frac{1}{6 \sqrt{6}}+\frac{9+4 \ln3}{12 \sqrt{6}}\epsilon.
\end{equation}
\subsection{NULL CIRCULAR ORBIT AND THE SHADOW}\label{Null CIRCULAR ORBITS}
In this subsection, we consider null circular geodesics in the AOS background. In this case, the effective potential can be obtained by setting $\alpha=0$ in \eqref{eq:effective}, while the expressions \eqref{eq:pt} and \eqref{eq:circulrOrbits} are still valid.
Solving $V(R)=0$ for $b\equiv L/E$, we obtain
\begin{equation}\label{eq:EL equation}
	b=\frac{R}{\sqrt{h(R)}}=\frac{R}{\sqrt{1-\left(\frac{2 m}{R}\right)^{\epsilon +1}}}\left(\frac{2 m}{R}\right)^{\epsilon }.
\end{equation}
 It should be noted that,
	$b$ is not strictly an impact parameter unless the metric component $h(r)\to 1$ at spatial infinity, which is clearly not the case here. However, it is a useful parameter for analyzing the motion of particles in the vicinity of the BH. Using the above relation together with the condition $V'(R)=0$, one quickly finds that the radius of the photon sphere can be obtained by solving the following equation
\begin{equation}\label{c2}
	2h(R)=Rh'(R).
\end{equation}
Comparing with  \eqref{c1}, it is easy to see that 
the circular null geodesic is actually the innermost circular time-like orbit \cite{Cardoso:2008bp}.
Thus we easily find out that the  radius of the photon sphere is $R_{\text{ph}}=R_{\text{ICO}}$ given by \eqref{eq:ICO}.  At this radius
\begin{equation}
	V''(R_{\text{ph}})=-\frac{\left(\epsilon-1\right)^3}{\epsilon-3}\frac{E^2}{m^2}
\end{equation}
which is negative for $\epsilon\ll1$. Thus, in AOS space-time, there is only one circular photon orbit at $r=R_{\text{ph}}$ 
and like Schwarzschild space-time, this orbit is unstable. At $r=R_{\text{ph}}$, the effective potential has the value
\begin{equation}
	V_{\text{max}}=\frac{E^2}{2R_{\text{ph}}^2}\left[b^2-b^2_{\text{ph}}\right]f(R_{\text{ph}})
\end{equation}
where $b_{\text{ph}}$ is obtained by inserting $R_{\text{ph}}$ into \eqref{eq:EL equation}
\begin{multline}\label{eq:impact}
	b_{\text{ph}}=2m\left[2+\frac{4}{\epsilon-3}\right]^{\frac{\epsilon-1}{\epsilon+1}}\sqrt{-1+\frac{4}{1+\epsilon}}\\ \approx 3\sqrt{3}m- 6 \sqrt{3} \left(\ln \frac{3}{2}\right) \left(\frac{m\gamma ^2 \Delta}{16 \pi }\right)^{1/3} 
\end{multline}
and in the second line, an $\epsilon$-expansion is used.  It is worth mentioning again that the first line of \eqref{eq:impact} is an exact result. 
From the second line, we see that the first-order quantum correction is proportional to $l_P^{2/3}$.
The ratio $b_{\text{ph}}/R_{\text{ph}}$ is given by
\begin{equation}
	\frac{b_{\text{ph}}}{R_\text{ph}}=\frac{\left(\frac{2 (\epsilon -1)}{\epsilon -3}\right)^{\frac{\epsilon }{\epsilon +1}}}{\sqrt{1-\frac{2 (\epsilon -1)}{\epsilon -3}}}\approx \sqrt{3}-\frac{2+3\ln \frac{3}{2}}{\sqrt{3}}\epsilon.
\end{equation}
It turns out that, as in the Schwarzschild case, this ratio is greater than one. For an initially ingoing photon with parameter $b$ that starts its motion at a sufficiently large value of $r$, there are three possibilities
according to \eqref{eq:eomI}:
\begin{itemize}
\item $\mathbf{\abs{b}<\abs{b_{\text{ph}}}}$. In this case $V_{\text{max}}<0$ and thus $\dot{r}^2$ is always a real positive quantity. Photons are eventually captured by the BH.
\item $\mathbf{\abs{b}=\abs{b_{\text{ph}}}}$. In this case $V_{\text{max}}=0$ and photons asymptotically reach $R_{\text{ph}}$ with an ever decreasing velocity $\dot{r}$. However, as mentioned above, this trajectory is unstable.
\item $\mathbf{\abs{b}>\abs{b_{\text{ph}}}}$. In this case $V_{\text{max}}>0$. Therefore, for some minimum value of $r$, the radial velocity $\dot{r}$ changes sign and the photons subsequently escape to infinity. Thus $b_{\text{ph}}$ determines the size of the BH shadow.
\end{itemize} 
According to \eqref{eq:impact}, the size of the AOS BH shadow decreases in comparison to its classical counterpart. This result is in agreement with the recent work \cite{Devi:2021ctm} in which the authors numerically studied the shadow of rotating and non-rotating AOS BHs. .Let us study this in more detail. Consider a static point source emitting photons in an  isotropic way and located at $r=r_0$ behind the AOS BH. 
The 4-vector tangent to its path is given by $u^{\mu}\varpropto \delta^{\mu}_{t}$ where the proportional constant can be obtained from $u^{\mu}u_{\mu}=-1$. The motion of the emitted photons is best analyzed in the local Fermi frame adapted to the source with the following basis
\begin{equation} \label{eq1}
	\begin{split}
		\bar{e}^{\mu}_t&=u^{\mu}=\frac{\delta^{\mu}_t}{\sqrt{h(r)}},\quad\quad \bar{e}^{\mu}_r=\sqrt{f(r)}\delta^{\mu}_{r}\\
		\bar{e}^{\mu}_{\theta}&=\frac{1}{r}\delta^{\mu}_{\theta},\quad\quad\quad\quad\quad \bar{e}^{\mu}_{\phi}=\frac{1}{r\sin\theta}\delta^{\mu}_{\phi}.
	\end{split}
\end{equation}
We will use $\{\bar{x}^{\mu}\}$ to denote Fermi-normal coordinates and $\{x^{\mu}\}$ will refer to an arbitrary coordinate system. The basis vectors $\bar{e}^{\mu}_{\alpha}$ form an orthonormal basis and then satisfy
$g_{\mu\nu}\bar{e}^{\mu}_{\alpha}\bar{e}^{\nu}_{\beta}=\eta_{\alpha\beta}$. Using Lagrangian \eqref{eq:first integral}, the canonical 4-momentum of the emitted photons  $p_{\mu}=\partial \mathcal{L}/\partial\dot{x}^{\mu}$  are given by
\begin{equation}
	p_{t}= -E,\quad p_{r}=\frac{\dot{r}}{f(r)},\quad p_{\phi}=L.
\end{equation}
if we assume that photons follow planar geodesics with $\theta=\pi/2$ and thus $p_{\theta}=0$. Using \eqref{eq1}, one can now find the four-momentum of the emitted photon in the  local Fermi frame. After some algebra we find
\begin{equation}\label{eq:momentums}
	\bar{p}^t=\frac{E}{\sqrt{h(r)}},\quad \bar{p}^r=\frac{\dot{r}}{\sqrt{f(r)}},\quad \bar{p}^{\phi}=\frac{L}{r}.
\end{equation}
Using \eqref{eq:momentums}, we obtain the velocity of the emitted photon measured in the local Fermi frame
\begin{equation}\label{v}
	\bar{v}^{r}=\frac{\bar{p}^{r}}{\bar{p}^t}=\frac{\dot{r}}{E},\quad \bar{v}^{\phi}=\frac{\bar{p}^{\phi}}{\bar{p}^t}=\frac{b}{r_0}\sqrt{h(r_0)}.
\end{equation}
Here, we are interested in the relation between the parameter $b$ and the emission angle $\eta$ defined as $v^x=\cos\eta$, $v^y=\sin\eta$ in which the Cartesian components of the radial velocity is used. Assuming that  photons are emitted at $\phi=0$, therefore, $\sin\eta=\bar{v}^{\phi}$ which is given by \eqref{v}.
The minimum angle for light rays to escape to infinity is obtained by substituting $b=b_{\text{ph}}$ from \eqref{eq:impact} into $\bar{v}^{\phi}$ in \eqref{v} to find the emission angle
\begin{multline}\label{eq:sin}
	\sin\eta_{min}=\frac{2 m}{r_0}\left(\frac{4}{\epsilon -3}+2\right)^{\frac{\epsilon -1}{\epsilon +1}} \sqrt{\frac{4}{\epsilon +1}-1}\cross\\
	\sqrt{\left(\frac{r_0}{2 m}\right)^{2 \epsilon } \left(1-\left(\frac{2 m}{r_0}\right)^{\epsilon +1}\right)}\\
	\approx\frac{3\sqrt{3}m}{r_0}\sqrt{1-\frac{2m}{r_0}}+\\
	3 \sqrt{3} \left(\frac{m\gamma ^2 \Delta}{16 \pi }\right)^{1/3}  \sqrt{1-\frac{2 m}{r_0}}
	\frac{m \ln\frac{8 r_0}{81 m}-r_0 \ln\frac{2 r_0}{9 m}}{r_0 \left(2 m-r_0\right)}
\end{multline}
 in which we  have also expanded the result in terms of $\epsilon$. This gives 
the effect of the quantum parameter on the escape angle.
The first term on the RHS of \eqref{eq:sin} is the well-known result of GR while the second term represents the correction due to the quantum effects. Since $\epsilon$ is extremely small, the escape angle in AOS space-time is not significantly different from that in Schwarzschild space-time. However, we want to know how this effect changes with the location of the source. Simple calculations show that $r_0\approx9.985m$ realizes the maximum of the quantum correction in \eqref{eq:sin} as:
\begin{equation}\label{emax}
	\sin\eta_{\text{min}}=0.465+0.464\epsilon
\end{equation}
\section{Geodetic Precession}\label{sec:Geodetic Precession}
In this section we assume that a gyroscope with spin $S$ is moving along a circular geodesic in the equatorial plane $\theta=\pi/2$ of AOS space-time. The 4-velocity of the gyroscope 
\begin{equation}
	u^{\mu}=\dot{t}\delta^{\mu}_t+\dot{\phi}\delta^{\mu}_\phi=\omega\left(\Omega^{-1}\delta^{\mu}_t+\delta^{\mu}_\phi\right)
\end{equation}
is orthogonal to its spin 4-vector $S^{\mu}$ along the geodesic: 
\begin{equation}\label{eq:su}
	S^{\mu}u_{\mu}=0.
\end{equation} 
Moreover, since the gyroscope is considered moving along a geodesic, its spin vector will be parallel transported along the orbit:
\begin{equation}\label{eq:parratra}
	\frac{dS^{\mu}}{d\tau}+\Gamma^{\mu}_{\nu\sigma}u^{\nu}S^{\sigma}=0.
\end{equation}
Using the connection coefficients of the metric \eqref{eq:metr}-\eqref{eq:f}, we notice that $S^{\theta}$ is constant. Therefore, we can set it to zero without loss of generality.  Moreover, the condition\eqref{eq:su} establishes a relation between $S^{t}$ and $S^{\phi}$ as follows
\begin{equation}
	S^{t}=\frac{R^2 \Omega} {h(R)}S^{\phi}.
\end{equation}
Therefore, we will consider only the dynamics of $S^{r}$ and $S^{\phi}$ which, according to \eqref{eq:parratra}, satisfy
\begin{equation}\label{eq:ode1}
	\frac{d}{d\tau}S^{r}(\tau)=-\mathcal{P} S^{\phi}(\tau),\quad \frac{d}{d\tau}S^{\phi}(\tau)=-\mathcal{Q}S^{r}(\tau)
\end{equation}
where
\begin{equation}
	\mathcal{P}\equiv \frac{R \omega }{2}\frac{R h'(R)-2 h(R)}{h(R)}f(R),\quad \mathcal{Q}\equiv \frac{\omega}{R}.
\end{equation}
In the above equations, we can convert the $\tau$-derivatives to $t$-derivatives by replacing $\omega$ by $\Omega$ in the definitions of $\mathcal{P}$ and $\mathcal{Q}$. It is often convenient to consider the local Fermi frame attached to the gyroscope. 
\begin{equation} \label{eq2}
	\begin{split}
		\bar{e}^{\mu}_{t}&=u^{\mu},\quad\quad\quad \bar{e}^{\mu}_r=\sqrt{f(R)}\delta^{\mu}_{R}\\
		\bar{e}^{\mu}_{\theta}&=\frac{1}{r}\delta^{\mu}_{\theta},\quad\quad\quad \bar{e}^{\mu}_{\phi}=\mathcal{X}\left[\Omega R \delta^{\mu}_{t}+h(R)\delta^{\mu}_{\phi}\right]
	\end{split}
\end{equation}
where
\begin{equation}
	\mathcal{X}=\frac{1}{R}\left[h(R) \left(h(R)-R^2 \Omega ^2\right)\right]^{-1/2}.
\end{equation}
It is easy to check that this basis is an orthonormal basis. The components of the spin of the gyroscope in the local Fermi frame are
\begin{equation} \label{eq3}
	\begin{split}
		\bar{S}^{t}&=0,\quad\quad\quad \bar{S}^{r}=\frac{S^{r}}{\sqrt{f(R)}}\\
		\bar{S}^{\theta}&=0,\quad\quad\quad \bar{S}^{\phi}=R \sqrt{1-\frac{R^2 \Omega ^2}{h(R)}}S^{\phi}
	\end{split}
\end{equation}
The evolution equations of the above components can be derived using \eqref{eq:ode1} and \eqref{eq3} as follows:
\begin{equation}\label{geod}
	\frac{d \bar{S}^{r}}{dt}=\Omega'\bar{S}^{\phi},\quad \frac{d\bar{S}^{\phi}}{dt}=-\Omega'\bar{S}^{r}
\end{equation}
where
\begin{multline} \label{x}
	\Omega'=\Omega  \sqrt{f(R)} \sqrt{1-\frac{R^2 \Omega ^2}{h(R)}}\\
	=\frac{1}{2R^2}\left(\frac{R}{2 m}\right)^{2 \epsilon }\left[2 \epsilon -(\epsilon -1) \left(\frac{2 m}{R}\right)^{\epsilon +1}\right]\cross\\
	\sqrt{1-\left(\frac{2 m}{R}\right)^{\epsilon +1}-\left[\frac{\left(\frac{R}{2 m}\right)^{\epsilon } \left(2 \epsilon -(\epsilon -1) \left(\frac{2 m}{R}\right)^{\epsilon +1}\right)}{2 R}\right]^2}
\end{multline}\\
is the rotation frequency of the spin vector in the comoving frame. According to the first line of the above equation, the spin frequency of the gyroscope in the comoving frame will vanish when the term under the square root in \eqref{x} is zero.
\begin{equation}
	h(R)=R^2\Omega^2.
\end{equation}
Substituting $\Omega^2$ from \eqref{eq:Kep} into the above equation, we obtain equation \eqref{c2} whose roots determine the locations of the photon spheres. From this we conclude that the spin frequency of the gyroscope in the comoving frame decreases as the particle tends toward the photon spheres. Therefore, $\Omega'$ vanishes once because, as discussed in Section \ref{Null CIRCULAR ORBITS}, there is only one photon sphere in AOS BH whose location is given by $R_{\text{ph}}=R_{\text{ICO}}$. To solve equations \eqref{geod}, we set the initial condition such that at $t=0$, the spin vector is in the radial direction i.e. $\bar{S}^{r}(0)=S_0$ and $\bar{S}^{\phi}(0)=0$, so that we obtain
\begin{equation}
	\bar{S}^{r}(t)=S_0\cos\Omega't,\quad \bar{S}^{\phi}(t)=-S_0\sin\Omega't.
\end{equation}
As mentioned before the effective metric \eqref{eq:metr} is asymptotically flat, therefore one can describe the precession of  the gyroscope with respect to the fixed stars with the frequency $\omega_{GP}=\Omega-\Omega'$, i.e. geodesic precession  frequency. In order to compare with the result of GR, let us expand $\omega_{GP}$ with respect to $\epsilon$:
\begin{multline}\label{ogp}
	\omega_{GP}\approx \omega_{GP}^{(GR)}+\omega_{GP}^{(\epsilon)}=\frac{3 m^{3/2}}{2 R^{5/2}}+\frac{5}{4 R^{3/2}}\left(\frac{\gamma ^2 \Delta }{16 \pi  \sqrt{m}}\right)^{1/3}
\end{multline}
where  we have expanded the quantum effect up to the leading order of $\epsilon$ i.e. $l_P^{2/3}$.
In the above expression, the $\omega_{GP}^{(GR)}$ is the well known result of GR and $\omega_{GP}^{(\epsilon)}$ shows  the LQG correction which is linear in $\epsilon$. For a gyroscope placed in a circular orbit about the Earth ($m_{\oplus}=4.4347\times10^{-3}\text{m}$), like the Gravity Probe B \cite{Everitt:2011hp,Everitt:2015qri} and using  $R_{\oplus}=(642+6371) \text{km}$, we find 
\begin{equation}\label{o over o}
\omega_{GP}^{(GR)}\approx 6.63  \text{arcsec}/\text{yr},\quad \omega_{GP}^{(\epsilon)}\approx 8.732 \cross 10^{-14}  \text{arcsec}/\text{yr}
\end{equation}
Since $\omega_{GP}^{(GR)}/\omega_{GP}^{(\epsilon)}\approx 7.6\times 10^{13}$ the quantum effects are too small to be detected at this scale. After one Kepler period, $T=2\pi/\Omega$, using the modified Kepler relation \eqref{Kepler Law}, the direction of spin changes by 
\begin{equation}\label{dphi}
	\Delta \phi_{rev}\approx\Delta\phi_{rev}^{(GR)}+\Delta\phi_{rev}^{(\epsilon)}\approx\frac{3 \pi  m}{R}+\pi \left(\frac{\gamma ^2 \Delta }{16 \pi  m^2}\right)^{1/3}
\end{equation}
in the same direction as the orbit. Evaluating the above relation for the Gravity Probe B setup, we get
\begin{equation}\label{phi1}
	\Delta\phi_{rev}^{(GR)}\approx1.2\cross10^{-3}\text{arcsec},\quad \Delta\phi_{rev}^{(\epsilon)}\approx6.4\cross10^{-18}\text{arcsec}.
\end{equation}
It is interesting to note that, up to the leading order, the term $\Delta\phi_{rev}^{(\epsilon)}$ is $R$-independent while $\Delta\phi_{rev}^{(GR)}$ is proportional to $R^{-1}$. This is a remarkable point since it means that as $R$ increases, $\Delta\phi_{rev}^{(GR)}$ diminishes but the contribution of the LQG correction remains constant. In other words,  according to \eqref{ogp}, $\omega_{GP}^{(GR)}$ decays faster than $\omega_{GP}^{(\epsilon)}$ and from \eqref{ogp} and \eqref{dphi} we get
\begin{equation}\label{ratios}
	\frac{\omega_{GP}^{(\epsilon)}}{\omega_{GP}^{(GR)}}\approx\frac{5R}{6m}\left(\frac{\gamma ^2 \Delta}{16 \pi  m^2 }\right)^{1/3}
	,\quad \frac{\Delta\phi_{rev}^{(\epsilon)}}{\Delta\phi_{rev}^{(GR)}}\approx\frac{R}{3m}\left(\frac{\gamma ^2 \Delta}{16 \pi  m^2 }\right)^{1/3}
\end{equation}
which means that the quantum effects become dominant at large enough scales. For example, for a gyroscope rotating around the Earth with the coordinate radius $R\approx 10^{17}\text{km}$, we have $\omega_{GP}^{(GR)}\sim \omega_{GP}^{(\epsilon)}$ and $\Delta\phi_{rev}^{(GR)}\sim\Delta\phi_{rev}^{(\epsilon)}$.  Another example is $Sgr A^*$ with $m\approx 10^6 M_\odot$\cite{GRAVITY:2021xju,Do:2019txf}, where $M_\odot$ is the solar mass. 
In this case, the quantum and GR effects are of the same order of magnitude at $R\approx 10^{36}\text{km}$. 

\section{CONCLUDING REMARKS}\label{se:CONCLUDING REMARKS}
In this paper the motion of free particles moving along circular time-like and null geodesics in AOS space-time has been investigated. Despite the complicated form of metric, it is interesting to note that all key relations in this paper such as $r_{\text{ICO}}$, $r_{\text{ISCO}}$, Kepler's law, etc have been reported with exact mathematical expressions. It was shown that, according to \eqref{eq:ICO} and\eqref{eq:RISCO} $r_{\text{ICO}}$ slightly increases compared to the Schwarzschild case while $r_{\text{ISCO}}$ decreases. However it is worth noting that these deviations are too small to be detected experimentally as one expected for an acceptable BH solution of LQG. Then we considered the effect of quantum parameter of metric on the shadow of the BH. Our result \eqref{eq:impact} shows that the size of shadow shrinks slightly  which is in agreement with a recent work on the subject \cite{Devi:2021ctm}. Moreover, we calculated the escape angle of photons emanating from a fixed source. The result shows that at some specific location of the source, the quantum effects on the the escape angle reach to its maximum value. However, due to the smallness of this effect, see \eqref{eq:sin}, it is practically impossible to observe it at present.

Finally, we turned our attention to the Geodesic Precession of a gyroscope which is free falling along a circular geodesic. Applying our results to the famous experimental setup, i.e. Gravity Probe B project, indicates that the geodesic precession frequency $\omega_{GP}^{(GR)}$ is about 13 orders of magnitude larger than its quantum modification $\omega_{GP}^{(\epsilon)}$,  see relation \eqref{o over o}. However, the interesting result of this paper is that at large enough scales such as those involved in the late universe evolution, the quantum effects surpasses GR effect. This can be clearly observed by looking at relation \eqref{dphi}. From this relation, one can see that the quantum effects introduce a constant value to the angle by which the spin of the gyroscope rotates after one Kepler period whereas the corresponding GR term falls-off as $1/R$. It should be noted that in such mentioned low curvature regime, both GR and quantum terms are much smaller than unity.  Although we do not expect to observe these effects experimentally, this result introduces a theoretical aspect of the AOS BH.
\begin{acknowledgments}
We would like to thank the anonymous referees for their valuable comments. R.S would like to thank Armin Sadeghi for his useful comments and suggestions. F. S is grateful to the University of Tehran for supporting this work under a grant provided by the university research council. 
\end{acknowledgments}
\appendix
\section{Paczynski-Wiita pseudo-Newtonian potential for the AOS black hole}\label{appen}
Pseudo-Newtonian potentials are potentials obtained by some modification of the Newtonian gravitational potential. 
These potentials have long been used in astrophysics. As a famous example, the Paczynski-Wiita pseudo-Newtonian potential \cite{Paczynski:1979rz} correctly gives the location of the ISCO in the spacetime containing a relativistic object such as a BH (for a review, see \cite{Abramowicz:2009bh}). For example, in \cite{Boonserm:2019nqq} the authors used this potential to obtain the location of the ISCO in Kottler spacetime. 
For metric \eqref{eq:metr}, the pseudo-Newtonian potential is 
\begin{align}\label{pn}
	\Phi(r)&=\frac{1}{2}\left(1-\frac{1}{h(r)}\right)\nonumber\\
	&=\frac{1}{2} \left(1-\left(\frac{2 m}{r}\right)^{2 \epsilon }\frac{1}{1-\left(\frac{2 m}{r}\right)^{\epsilon +1}}\right)
\end{align}
As mentioned in section \ref{sec:TIME-LIKE CIRCULAR ORBITS}, according to \cite{Faraoni:2020stz} in the AOS background, the calculations with the $\epsilon$-expanded Paczynski-Wiita pseudo-Newtonian potential lead to a divergent quantum correction of the position of the ISCO. Here we use the exact form of \eqref{pn} to show that this is not the case, and recover \eqref{isco} , which gives no divergence in the quantum correction. 
The effective potential is obtained by adding the angular momentum $L$ as \cite{Abramowicz:2009bh} 
\begin{equation}
	V_{\text{eff}}(r)=\Phi(r)+\frac{L^2}{2r^2}
\end{equation}
with
\begin{multline}
	V_{\text{eff}}'(r)=\\
\frac{\left(\frac{2 m}{r}\right)^{2 \epsilon }}{2 r \left(1-\left(\frac{2 m}{r}\right)^{\epsilon +1}\right)}\Bigg[2\epsilon+	\frac{(\epsilon +1) \left(\frac{2 m}{r}\right)^{\epsilon +1}}{1-\left(\frac{2 m}{r}\right)^{\epsilon +1}}\Bigg]-\frac{L^2}{r^3}
\end{multline}
and
\begin{multline}\label{pnVpp}
	V_{\text{eff}}''(r)= \frac{\left(\frac{2 m}{r}\right)^{2 \epsilon }}{2 r^2 \left(1-\left(\frac{2 m}{r}\right)^{\epsilon +1}\right)}\Bigg[-2 \epsilon  (2 \epsilon +1)+\\
	\frac{(\epsilon +1) \left(\frac{2 m}{r}\right)^{\epsilon +1} \left(3 \epsilon  \left(\frac{2 m}{r}\right)^{\epsilon +1}-(5 \epsilon +2)\right)}{\left(1-\left(\frac{2 m}{r}\right)^{\epsilon +1}\right)^2}\Bigg]+\frac{3 L^2}{r^4}
\end{multline}
The angular momentum is obtained by solving $V_{\text{eff}}'(R)=0$ as 
\begin{equation}
	L^2=-R^2 \left(\frac{2 m}{R}\right)^{2 \epsilon }\frac{(\epsilon -1) \left(\frac{2 m}{R}\right)^{\epsilon +1}-2 \epsilon }{2 \left(1-\left(\frac{2 m}{R}\right)^{\epsilon +1}\right)^2}
\end{equation}
where $R$ is the radius of the orbit \cite{Abramowicz:2009bh,Faraoni:2015lud}.
Substituting the above relation into \eqref{pnVpp} we get 
\begin{multline}
	V_{\text{eff}}''(R)=-\frac{\left(\frac{2 m}{R}\right)^{2 \epsilon }}{2 R^2 (1-x)^3}\Bigg[x^2\left(\epsilon ^2-4 \epsilon +3\right)-\\
	x\left(3 \epsilon ^2-12 \epsilon +1\right)+4 (\epsilon -1) \epsilon\Bigg]
\end{multline}
where $x\equiv \left(2m/R\right)^{1+\epsilon}$. The possible locations of $R_{ISCO}$ are the roots of $V_{\text{eff}}''(R)=0$\cite{Boonserm:2019nqq}\footnote{One can also solve $dL/dR=0$ to obtain the ISCO radius. See \cite{Paczynski:1979rz} for more discussion.}. This reduces to 
\begin{equation}
	x^2\left(\epsilon ^2-4 \epsilon +3\right)-
	x\left(3 \epsilon ^2-12 \epsilon +1\right)+4 (\epsilon -1) \epsilon=0.
\end{equation}

It is easy to see that the above equation and equation \eqref{quad} are the same. Therefore, we end up with \eqref{isco} for $R_{\text{ISCO}}$. As expected, the Paczynski-Wiita pseudo-Newtonian approach and the GR approach give the same result for the ISCO radius. Note that this quantity is finite in the limit of $\epsilon\to 0$.

\end{document}